\documentclass{ws-mpla}
\usepackage[super]{cite}
\usepackage{graphicx}
\usepackage{xcolor}

\begin{document}

\markboth{Ciprian A. Sporea}
{Scattering of quantum fields by a MOG Black Hole}

\catchline{}{}{}{}{}

\title{Scattering of quantum fields by a MOG Black Hole }

\author{\footnotesize CIPRIAN A. SPOREA}

\address{Faculty of Physics, West University of Timi\c soara,\\
V.  P\^ arvan Ave.  4, RO-300223 Timi\c soara, Romania\\
ciprian.sporea@e-uvt.ro}

\maketitle

\pub{Received (Day Month Year)}{Revised (Day Month Year)}

\begin{abstract}
This paper aims to investigate the scattering of fermions by spherically symmetric MOG black holes, that are a type of black holes encountered in scalar-tensor-vector modified gravitational theories. After determining the scattering modes in this black hole geometry, we apply the partial wave method to compute analytical expressions for the phase shifts that enter into the definition of scattering amplitudes. An analysis of the influence of the MOG parameter $\alpha$ on the differential scattering cross section and the induced polarization is conducted. Also, a comparison with Schwarzschild scattering (for which $\alpha = 0$) is performed. Furthermore, it is also shown that glory and spiral/orbiting scattering are more significant for higher values of the free parameter $\alpha$.

\end{abstract}

\ccode{PACS Nos.:04.70.-s, 04.62.+v.}

\section{Introduction}

The latest astrophysical measurements on the composition of the Universe \cite{Plank2015} reveal that it consists of only 4.9$\%$ ordinary matter and energy, $26.8\%$ dark matter and the rest of $68.3\%$ of the so called dark energy, that we know nothing about. If dark matter is made of weakly interacting particles, that interact only gravitationally with visible parts of our universe, they have thus far failed to be detected \cite{Bertone,DM}. Some indirect evidences for the existence of dark matter come, for example, from the missing mass that needs to be added to galaxies in order to explain the flattens of galaxy rotation curves \cite{Rubin1,Rubin2,Sofue}. However, in the last two decades there are studies that succeed to explain the flatness of galaxy rotation curves \cite{milig,sanders,moffat,capozz1} and other problems without the need for existence of dark matter, but by proposing different models of modified theories of gravity (see, for example, the reviews \cite{Clifton,Capozzielo}).

The existence of black hole solutions in modified theories of gravity was confirmed by now in numerous studies in the literature (see, for example, Refs. \cite{bhsol1,bhsol2,bhsol3} and the references therein). Here we will focus our attention on a black hole solution coming from a covariant scalar-tensor-vector modified gravitational theory (MOG for short), theory first proposed by Moffat in Ref. \cite{MOG1}. This particular modified gravity theory was used with some success in explaining  the flatness of galaxy rotation curves, the dynamics of galaxy clusters, the matter power spectrum, the CMB acoustical power spectrum and also other physical features \cite{MOG2,MOG4,MOG5,MOG6,MOG7,MOG8}.

Furthermore, only recently the existence of classical spherically symmetric Schwarzschild-MOG black holes and rotating Kerr-MOG black hole solutions in MOG theory were found \cite{MOG9,MOG10}. Since then its thermodynamics properties were investigated \cite{Moffat1}, the existence of stable circular orbits \cite{Lee-Han}, accretion disks \cite{Perez}, quasinormal modes \cite{moffat2} or superradiance \cite{MOG4a} are just a few examples.

Moreover, up to our knowledge this study is the first one in the literature on the problem of fermion scattering by MOG black holes. In studying this problem we will use the methods already developed in our previous papers  \cite{sporea1,sporea2}, that basically consist in finding scattering modes solutions to the Dirac equation in the MOG black hole geometry and then use a partial wave method on these modes in order to obtain analytical expressions for the phase shifts that enter into the definitions of scattering cross sections.

Regarding other studies that exist in the literature dedicated to the problem of quantum fields scattering by different types of black holes, one can find for example that absorption and scattering of massive and massless scalar waves are investigated in Refs. \cite{futterman,sancez1,sancez2,jung,crispino1,crispino2,crispino3,crispino4,andersson,dnpage,gaina,gainaa,matzner1}. Half-integer spin waves, associated to fermions, were less studied compared with the other types of integer-spin waves. But studies on fermions scattered by Schwarzschild BHs. can be found in Refs. \cite{sporea1,unruh,dolan,dolan1,das,jin,doran,park,cho-lin,sporea3,sporea6}, on charged Reissner-Nordstrom BHs. in Refs. \cite{sporea2,sporea3,gaina2}, on Kerr BHs. in Refs. \cite{gaina1,Lee} or regular BHs. in Refs. \cite{huang,sporea5}. Furthermore, in Ref. \cite{CrispinoDolan} one can find a summary of Schwarzschild scattering for fields with spin $s=0,\,1/2,\,1,\,2$. For scattering of electromagnetic and gravitational waves by different types of black hole spacetimes one can consult for ex. Refs. \cite{crispino5,crispino6,dolan3,crispino7,crispino8,mashhoon1,mashhoon2,fabbri,logi,matzner2,matzner3}.

The remaining of this paper is structured as follows: in Section 2 we start by introducing the MOG black hole metric and continue with a brief review of the Dirac equation in black hole spacetimes. Section 3 is dedicated to presenting the scattering modes obtained by solving the Dirac equation in the asymptotic part of the black hole geometry, followed by the computation of the phase shifts that will allow us in Section 4 to discuss the main results by analyzing the differential scattering cross section and the induced polarization. The last Section 5, contains the conclusions of the paper. In this paper we use natural units with $c=\hbar=G_N=1$.

\section{Main equations}\label{sec.2}

The solution describing a black hole geometry in modified gravity (MOG) theory was first obtained in Refs. \cite{MOG9,MOG10}, where it was showed that the spherically symmetric MOG black hole has the following metric tensor
\begin{equation}\label{mogbh}
\begin{split}
& ds^2 =h(r)dt^2 - \frac{dr^2}{h(r)} -r^2\left( d\theta^2+\sin^2\theta d\phi^2 \right), \\
& h(r) = 1-\frac{2(1+\alpha)MG_N}{r} +\frac{\alpha(1+\alpha)M^2G_N^2}{r^2} ,
\end{split}
\end{equation}
where $M$ is the mass of the black hole, $G_N$ is Newton's constant and $\alpha$ is a free parameter defined by an enhanced gravitational constant $G=(1+\alpha)G_N$. Furthermore, we observe that the MOG black hole solution (\ref{mogbh}) has the form of the Reissner-Nordstr\"om solution for an electrically charged black hole. This feature makes possible to solve the Dirac equation as in the geometry of a Reissner-Nordstr\"om black hole \cite{sporea2}.

The metric tensor (\ref{mogbh}) results from solving the following field equations \cite{MOG9}
\begin{equation}\label{mog2}
\begin{split}
& R_{\mu\nu}-\frac{1}{2}g_{\mu\nu}R=-8\pi G T^{(\phi)}_{\mu\nu} , \\
& \frac{1}{\sqrt{-g}}\partial_\nu\left( \sqrt{-g}B^{\mu\nu} \right)=0 ,\\
& \partial_\sigma B_{\mu\nu} +\partial_\mu B_{\nu\sigma} +\partial_\nu B_{\sigma\mu} =0 .
\end{split}
\end{equation}
In scalar-tensor-vector gravity theory (STVG) \cite{MOG1} $B_{\mu\nu}=\partial_\mu\phi_\nu-\partial_\nu\phi_\mu$, represents the field strength of a vector field $\phi_\mu$ associated to a repulsive gravitational force with the source charge $Q_g=M\sqrt{\alpha G_N}$. This charge originates from the definition \cite{Moffat1} $Q_g=k\int d^3xJ^0(x)$, where $J^0(x)$ is the 0-th component of the covariant current density $J^\mu=kT^{(M)}_{\mu\nu} u_\nu$, with $T^{(M)}_{\mu\nu}$ the energy-momentum tensor of ordinary matter and $u^\mu=dx^\mu/ds$, $s$ being the proper time. The values of $Q_g=M\sqrt{\alpha G_N}$ and $G=(1+\alpha)G_N$ are obtained from the weak field approximation of the MOG field equations \cite{MOG1,MOG2}.

 As in Einstein gravity the matter energy-momentum tensor was set to zero $T^{(M)}_{\mu\nu}=0$, while the energy-momentum tensor of the vector filed $\phi_\mu$ is given by
\begin{equation}
T^{(\phi)}_{\mu\nu} = -\frac{1}{4\pi}\left( B_\mu^{\,\,\,\alpha}B_{\nu\alpha} -\frac{1}{4}g_{\mu\nu}B^{\alpha\beta}B_{\alpha\beta} \right) .
\end{equation}
The horizon structure of the MOG black hole (\ref{mogbh}) is the same as in the case of Reissner-Nordstr\"om electrically charged black hole, having two horizons
\begin{equation}
r_\pm=MG_N\left[ 1+\alpha \pm \sqrt{1+\alpha} \right],
\end{equation}
with $r_+$ the radius of the black hole event horizon and $r_-$ an inner Cauchy horizon. By setting $\alpha=0$, the metric (\ref{mogbh}) reduces to the usual general relativity Schwarzschild metric and $r_+=r_s=2G_N M$. By imposing the conditions $r_+\geq0$ and $Q_g\geq0$, we obtain the range of the parameter $0\leq\alpha<\infty$. In week gravitational fields, the resulted modified Newtonian acceleration law  can be fitted against galaxy rotation curves data. The value of $\alpha$ resulted from these fits is \cite{MOG2}: $\alpha=8.89$. Also in Ref. \cite{MOG1} it was shown that Mercury's perihelion advance is in agreement with the GR result for $\alpha<<1$. However, in strong gravitational fields, the value of $\alpha$ resulted from the week field determination, is no longer valid \cite{Moffat1}. Thus, for MOG black holes we will treat $\alpha$ as a free parameter.

In the remaining of this section we will briefly review the Dirac equation and deduce its expression in the geometry of a MOG black hole. We consider here the case in which the Dirac equation does not couple to the vector field $\phi_\mu$ and couples only to the spacetime geometry.

As was showed in Refs. \cite{cota,cota1} the Dirac equation, $ i\gamma^a D_a\psi -m\psi = 0$, for a black hole spacetime with spherical symmetry can be put into a Hamiltonian form $H_D\psi(x)=i\partial_t\psi(x)$, with
\begin{equation}
H_D= -i\frac{h(r)}{r^2}(\gamma^0\gamma^i x^i)\left( 1+
x^i\partial_i\right)-i\frac{\sqrt{h(r)}}{r^2}(\gamma^i x^i){\gamma^0 \left(2\vec{S}\cdot \vec{L} +1\right)}+\sqrt{h(r)}\gamma^0 m .
\end{equation}
In obtaining the above equation the Cartesian gauge \cite{Vilalba,cota1} was used and in doing so the resulting angular part of the Dirac equation is the same as on flat spacetimes, such that now the particle-like solution can be described by the wave function
\begin{equation}\label{rad1}
\begin{split}
\psi(x)=&\psi_{E,j,m,\kappa}(t,r,\theta,\phi) =\\
&\frac{e^{-iEt}}{r\,h(r)^{1/4}}\left\{ f^+_{E,\kappa}(r)\Phi^+_{m,\kappa}(\theta, \phi) + f^-_{E,\kappa}(r)\Phi^-_{m,\kappa}(\theta, \phi) \right\} ,
\end{split}
\end{equation}
where $\Phi^\pm_{m,\kappa}(\theta, \phi)$ are the usual 4-component angular spinors \cite{Thaller,Landau} that are orthogonal to each other. The spinors are completely determined by the energy $E$, the overall angular momentum $j$, the spin quantum number $m=\pm1/2$ and the eigenvalue $\kappa$ of the spin-orbit operator. We are using for $\kappa$ the convention $\kappa=\pm(j+1/2)$ with $l=|\kappa|-(1-sign\,\kappa)/2$, where $l$ is the orbital angular momentum.

After taking care of the angular part of the Dirac equation, one remains only with a radial Dirac equation that needs to be solved. The equation containing the unknown radial wave functions $f^\pm_{E,\kappa}(r)$ turns out to be \cite{sporea1,sporea4}:
\begin{equation}\label{rad6}
\renewcommand{\arraystretch}{1.8}
\left(\begin{array}{cc}
m\,\sqrt{h(r)}-E& -h(r)\frac{\textstyle d}{\textstyle dr}+\frac{\textstyle \kappa}{\textstyle r}\sqrt{h(r)}\\
h(r)\frac{\textstyle d}{\textstyle dr}+\frac{\textstyle \kappa}{\textstyle r}\sqrt{h(r)}& -m\,\sqrt{h(r)}-E
\end{array}\right)\left(\begin{array}{c}
f^+_{E,\kappa}(r)  \\
f^-_{E,\kappa}(r)
\end{array}\right) = 0 .
\end{equation}
Making in eq. (\ref{rad6}) the substitution of the function $h(r)$, taken from eq. (\ref{mogbh}), we obtain the form of the radial Dirac equation in the particular case of a MOG black hole spacetime.

\section{Scattering modes, phase shifts and scattering cross sections}\label{sec.scatt}

For a complete solution of the scattering problem, Eq. (\ref{rad6}) has to be solved in its full range. Unfortunately, the radial equations (\ref{rad6}) can not be solved analytically in their present form. However, asymptotic scattering modes can be obtained by solving the radial equation (\ref{rad6}) in the asymptotic part of the modified gravity (MOG) black hole geometry using the same procedure as in our previous papers \cite{sporea1,sporea2} on Schwarzschild and Reissner-Nordstrom black hole scattering. For this reason, here we will resume to only give the main steps for deriving the scattering modes solution.

One starts with the introduction of a new variable $x$ defined by
\begin{equation}\label{x}
x=\sqrt{\frac{r}{r_{+}}-1}\,\in\,(0,\infty), \qquad r_+=MG_N\left[ 1+\alpha + \sqrt{1+\alpha} \right],
\end{equation}
followed by the multiplication with $(1+x^2)/x$ of eq. (\ref{rad6}), that will produce a new set of equations for the radial wave functions  $f^\pm_{E,\kappa}(x)$. Because we are interested in the scattering modes, we can use a Taylor expansion of the equations with respect to $1/x$ (neglecting the $O(x^2)$ terms and of higher order) to obtain in the end the new approximative equations
\begin{equation}\label{sol7}
\openup 2\jot
\begin{split}
& \left( \frac{1}{2}\frac{d}{dx}+\frac{\kappa}{x} \right)f^+_{E,\kappa}(x) - x(\varepsilon+\mu)f^-_{E,\kappa}(x)  - \frac{1}{x}\left(\varepsilon+ \zeta \right)f^-_{E,\kappa}(x) = 0,  \\
& \left( \frac{1}{2}\frac{d}{dx}-\frac{\kappa}{x} \right)f^-_{E,\kappa}(x) + x(\varepsilon-\mu)f^+_{E,\kappa}(x)  + \frac{1}{x}\left(\varepsilon - \zeta \right)f^+_{E,\kappa}(x) = 0,
\end{split}
\end{equation}
where the following notations were used
\begin{equation}
\mu=r_{+}m\,,\quad \varepsilon=r_{+}E\,,\quad \zeta=\frac{1}{2}\mu\left(1-\frac{r_-}{r_+}\right).
\end{equation}
By introducing the functions $\hat f^\pm_{E,\kappa}(x)$ defined by
\begin{equation}\label{mogbh1}
\left(\begin{array}{c}
f^+_{E,\kappa}(x)  \\
f^-_{E,\kappa}(x)
\end{array}\right)=
\left(
\begin{array}{c}i\sqrt{\varepsilon+\mu}\,(\hat f^- -\hat f^+)\\
\sqrt{\varepsilon-\mu}\,(\hat f^+ +\hat f^-)
\end{array}\right),
\end{equation}
the new equations that will result, can be solved analytically in terms of Whittaker $M$ and $W$ special functions, to obtain \cite{sporea1,sporea2,cota}:
\begin{equation}\label{scatmodes}
\begin{split}
& \hat f^+(x)=C_1^+\frac{1}{x}M_{\rho_+,s}(z)+C_2^+\frac{1}{x}W_{\rho_+,s}(z) \\
& \hat f^-(x)=C_1^-\frac{1}{x}M_{\rho_-,s}(z)+C_2^-\frac{1}{x}W_{\rho_-,s}(z),
\end{split}
\end{equation}
where $z=2i\nu x^2\,$, $\nu=\sqrt{\varepsilon^2-\mu^2}$ and the following notations were introduced
\begin{equation}
s=\sqrt{\kappa^2+\zeta^2-\varepsilon^2},\quad
\rho_{\pm}=\mp\frac{1}{2}-i q,\quad q= \frac{\varepsilon^2-\zeta\mu}{\nu} \,.
\end{equation}
Furthermore, the integration constants satisfy the relations \cite{cota,sporea1}
\begin{equation}\label{C1C1}
\frac{C_1^-}{C_1^+}=\frac{s-i q}{\kappa-i\lambda}\,,\quad
\frac{C_2^-}{C_2^+}=-\frac{1}{\kappa-i\lambda}\,,\quad \lambda=\frac{\varepsilon\left(\mu-\zeta\right)}{\nu}\,.
\end{equation}
Applying the partial wave analysis on the scattering modes (\ref{mogbh1}), after imposing the condition $C^\pm_2=0$ in order to recover the correct newtonian limit for large values of the angular momenta $l$ (more details can be found in Appendix C from Ref. \cite{sporea1}), the following point-independent phase shifts are obtained
\begin{equation}\label{phase}
e^{2i\delta_{\kappa}}=\frac{\kappa-i\lambda}{s-iq}\cdot\frac{\Gamma(1+s-iq)}{\Gamma(1+s+iq)}\cdot e^{i\pi(l-s)},
\end{equation}
with the help of which the following scattering amplitudes can be computed
\begin{equation}\label{sum1}
\begin{split}
&f(\theta)=\sum_{l=0}^{\infty}\frac{1}{2ip}\left[ (l+1)(e^{2i\delta_{-l-1}}-1)+ l(e^{2i\delta_l}-1)\right]\,P_l^0(\cos \theta), \\
&g(\theta)=\sum_{l=1}^{\infty}\frac{1}{2ip}\left[e^{2i\delta_{-l-1}}-e^{2i\delta_l}  \right]\,P_l^1(\cos\theta).
\end{split}
\end{equation}
The differential scattering cross section is obtained as the sum of squares of the two scalar amplitudes, namely
\begin{equation}
\frac{d\sigma}{d\Omega}=|f(\theta)|^2 + |g(\theta)|^2.
\end{equation}
Although, the above phase shifts (\ref{phase}) have the same formal expressions as in the case of fermion scattering by Schwarzschild \cite{sporea1} and Reissner-Nordstrom \cite{sporea2} black holes, the parameters involved ($s, q, \lambda$) have a different dependence on the relevant physical parameters $m$ and $E$ (fermion's mass and energy), $v$ (velocity) and $M$ (black hole mass):
\begin{equation}
\begin{split}
& s^2=\kappa^2 + (mM)^2(1+\alpha)-(ME)^2\left[ 1+\alpha+\sqrt{1+\alpha} \right]^2, \\
& q = vME\left[ 1+\alpha+\sqrt{1+\alpha} \right] + \frac{m^2M}{vE},\qquad \lambda=\frac{mM}{v},
\end{split}
\end{equation}
resulting in important differences in the scattering cross sections as can be seen in Section \ref{sec.results}. The series in Eq. (\ref{sum1}) contain a singularity in $\theta=0$ and as a consequence the series are poorly convergent. This problem can be dealt with by using the method of reduced series first proposed in ref. \cite{Yennie} and later on used in Refs. \cite{dolan,sporea1,sporea2} and extended to gravitational waves in Ref. \cite{mashhoon2}. The plots presented in the next section are obtained using the reduced series as given in eqs. (72)-(75) from Ref. \cite{sporea1}.

\section{Main results and discussions}\label{sec.results}

For a better understanding of the analytical results obtained in the previous section, in what follows we will make a series of plots of the differential scattering cross section and the polarization degree induced by the scattering of fermions by MOG black holes in modified gravity (MOG) theory. For computing the polarization the following formula is used \cite{rose}
\begin{equation}\label{pwa35}
\mathcal{P}=-i\frac{fg^*-f^*g}{|f|^2 + |g|^2}.
\end{equation}

Despite of the fact that the MOG black hole can be mapped into a Reissner-Nordstr\"om black hole, when writing the Dirac equation for the MOG black hole the term containing the electrostatic potential $eQ/r$ is missing and as a consequence the factors of the type $eQ$ disappear from the parameters involved in the scattering modes. This also implies that we do not have the analogue of $eQ$ present in the MOG black hole phase shifts. Thus, if in our previous paper \cite{sporea2} we analysed the scattering of fermions by a Reissner-Nordstr\"om black hole using especially the parameter $eQ$, here we can not do this, and what is new in this paper is that the analysis is done with the MOG parameter $\alpha$, that is not the equivalent of $eQ$. Thus, the results differ from those in Ref. \cite{sporea2}, although the same phenomena of glory and spiral/orbiting scattering can be observed for both types of black holes.

Furthermore, a neutral massless Dirac field on Reissner-Nordstr\"om also does not have $eQ/r$ terms and using the same mapping of the metric the case of a neutral Dirac field on Reissner-Nordstr\"om is obtained. This scenario was studied in our previous paper \cite{sporea3}.

\begin{figure*}
	\centering
	\includegraphics[scale=0.30]{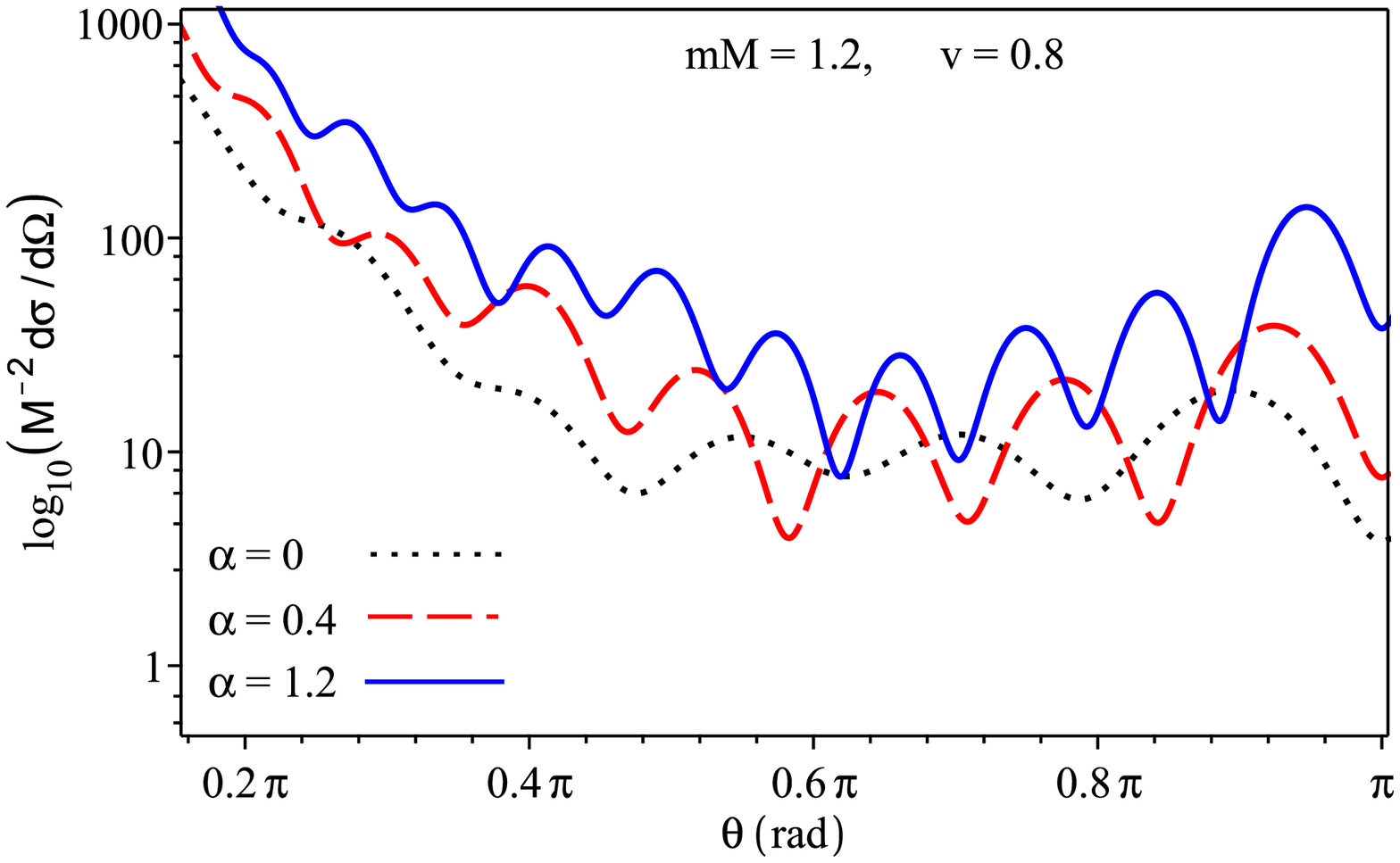}
	\includegraphics[scale=0.30]{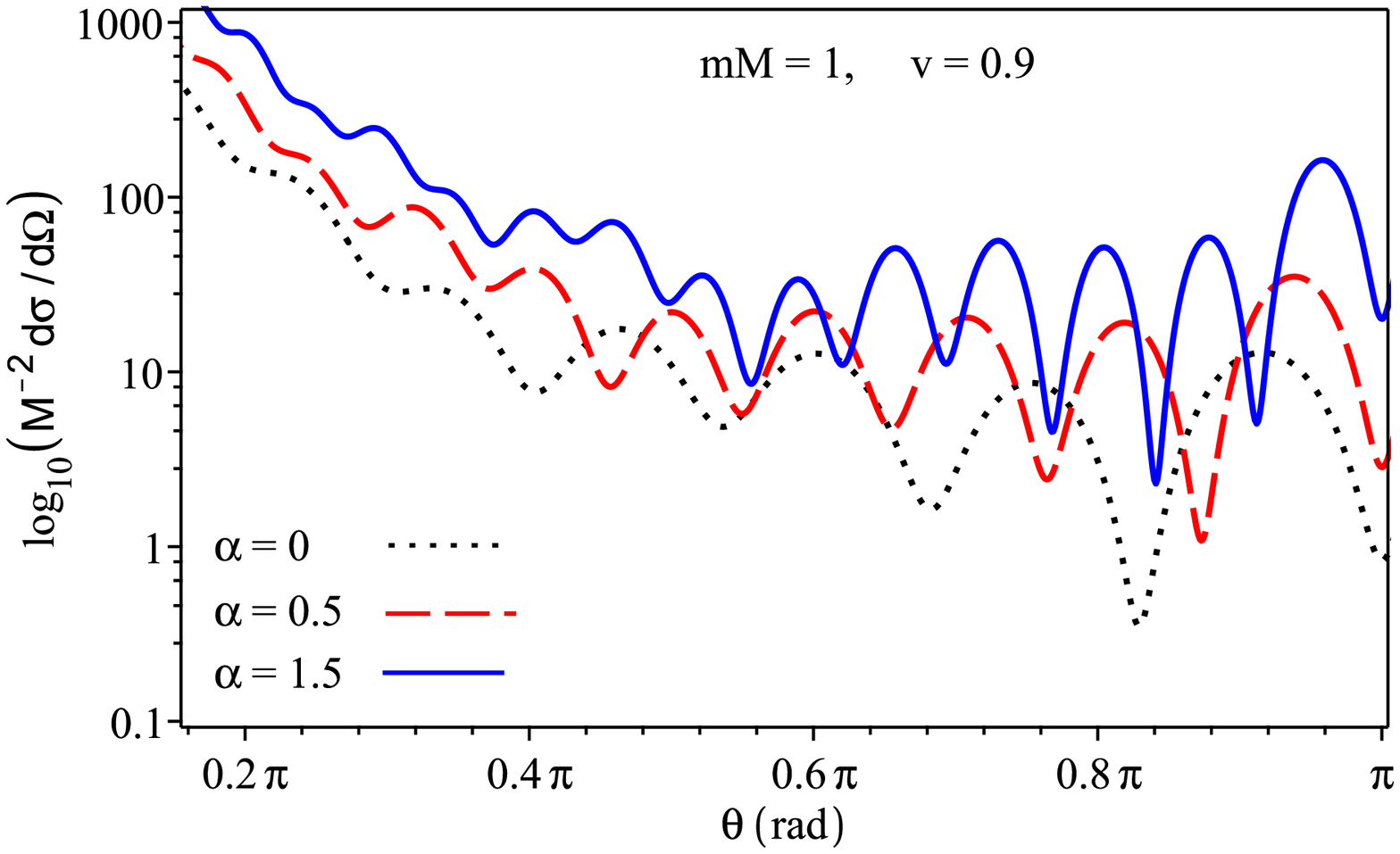}
	\includegraphics[scale=0.3]{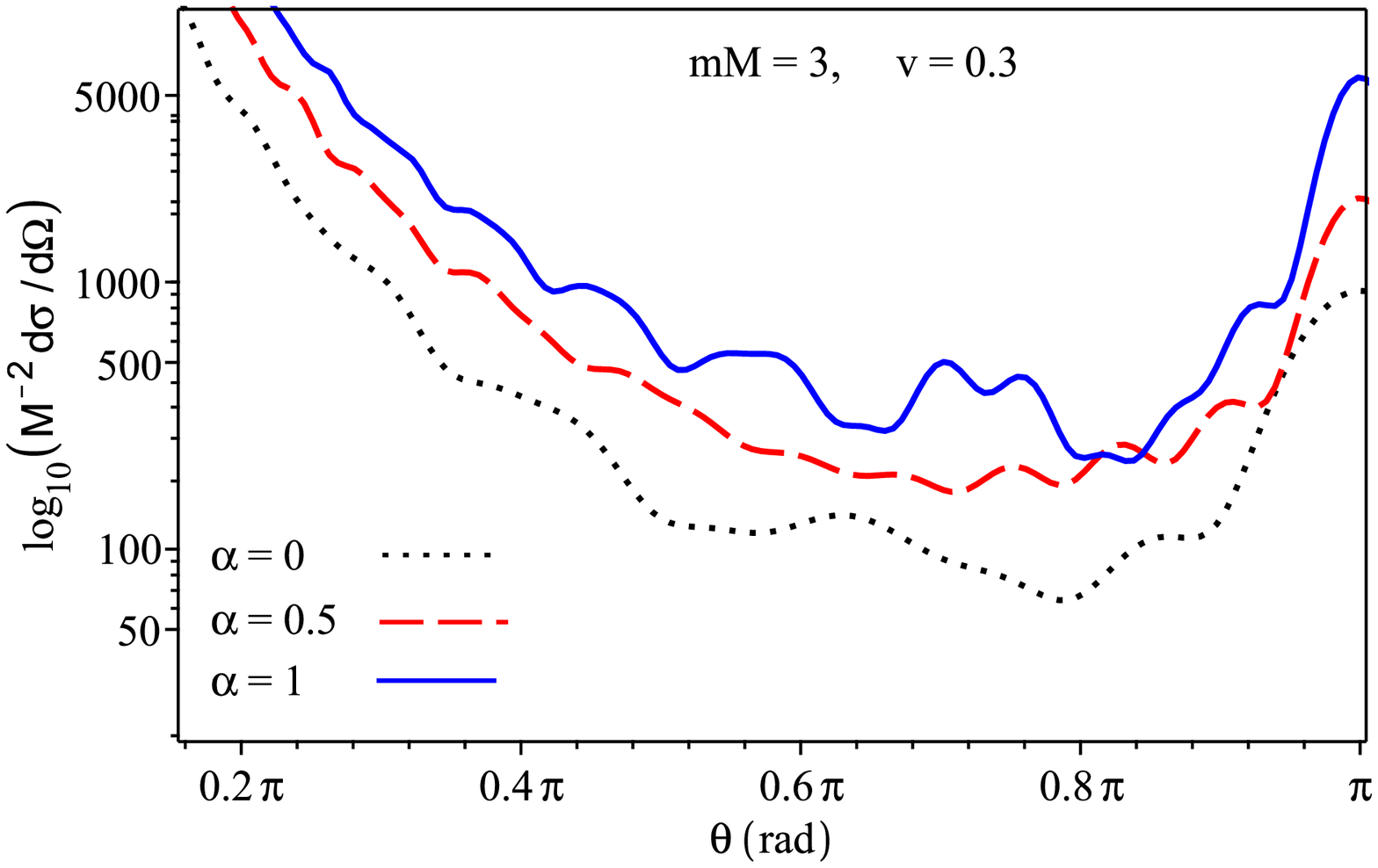}
	\includegraphics[scale=0.3]{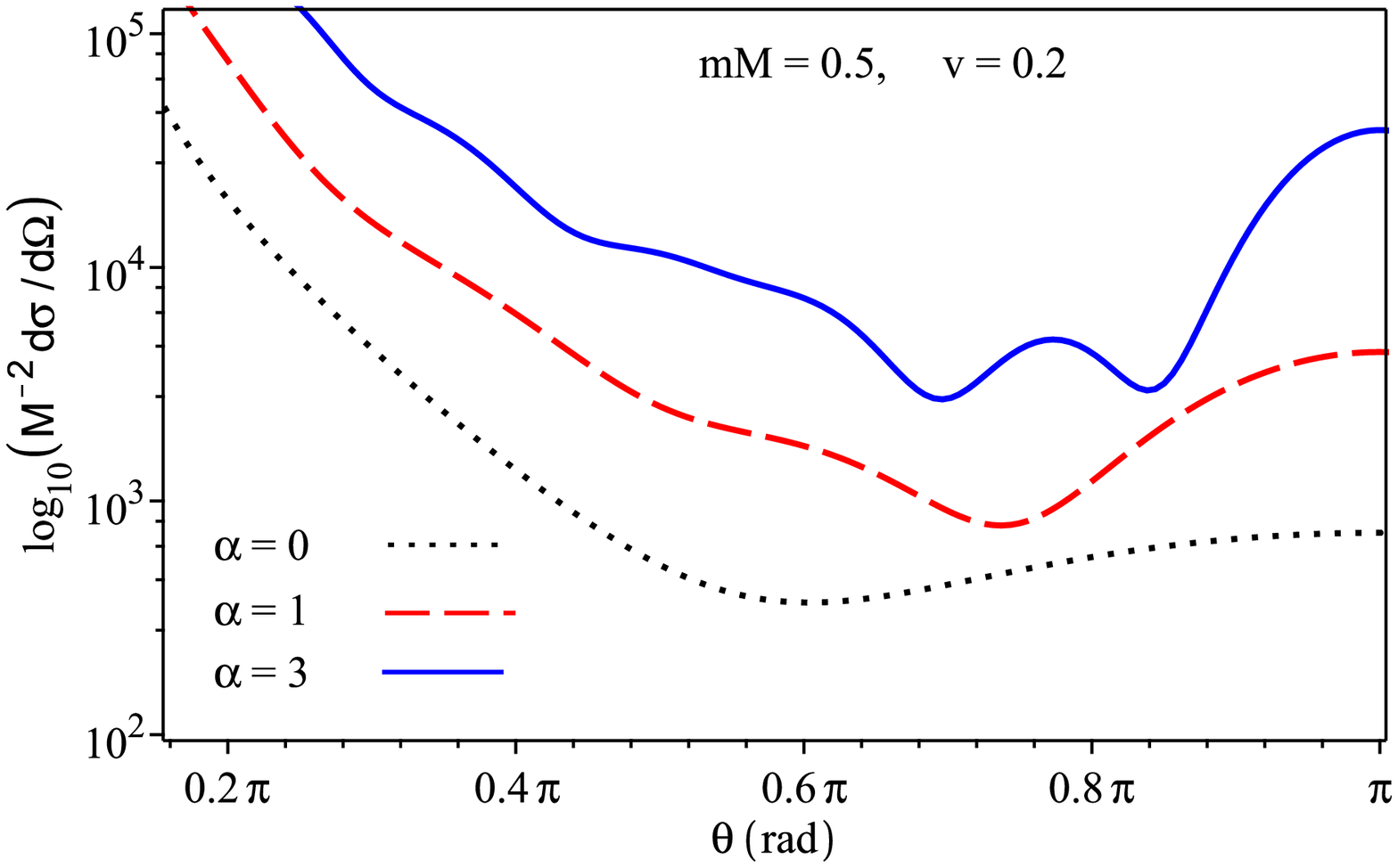}
	\caption{(color online). Differential scattering cross section as a function of the scattering angle $\theta$ in the case of scattering of relativistic (top panels) and non-relativistic (bottom panels) fermions by a MOG black hole. Both glory and spiral/orbiting scattering can be observed. The curves with $\alpha=0$ correspond to Schwarzschild black holes. }
	\label{fig1}
\end{figure*}

Analyzing the influence of the modified gravity (MOG) parameter $\alpha$ on the differential scattering cross section (see Fig. \ref{fig1}), one can observe an enhancement in the scattering intensity as the value of $\alpha$ is increased. Since the free parameter $\alpha$ defines the enhanced gravitational constant \cite{MOG1,MOG2} $G=(1+\alpha)G_N$, we can assume that a black hole with a higher value of $\alpha$, will produce a stronger gravitational field, that will increase the probability for the particle to be scattered at any angle. This is valid both for scattering of relativistic fermions (upper panels in Fig. \ref{fig1}) and also for scattering of non-relativistic fermions (bottom panels in Fig. \ref{fig1}). We use the term non-relativistic for fermions with velocities smaller than $v=0.4c$ (with $c=1$ in natural units) and the term relativistic for fermions with velocities bigger than $v=0.7c$. Furthermore, the glory scattering (associated to the existence of a halo or a bright spot in the backward direction $\theta=\pi$) becomes more pronounced as $\alpha$ takes higher values. As in the case of fermion scattering by Schwarzschild black holes \cite{dolan,sporea1}, one can also have spiral (or orbiting) scattering for MOG black holes. Spiral scattering is visualised by the presence of oscillations in the scattering intensity. From Fig. \ref{fig1} one can also observe that the oscillations present in the scattering intensity become more frequent as the value of $\alpha$ is increased, which means that the spiral scattering is more significant for MOG black holes compared with Schwarzschild black holes (that have $\alpha=0$). Another thing to point out is the fact that spiral scattering can be better observed for scattering of relativistic fermions.

\begin{figure*}
	\centering
	\includegraphics[scale=0.3]{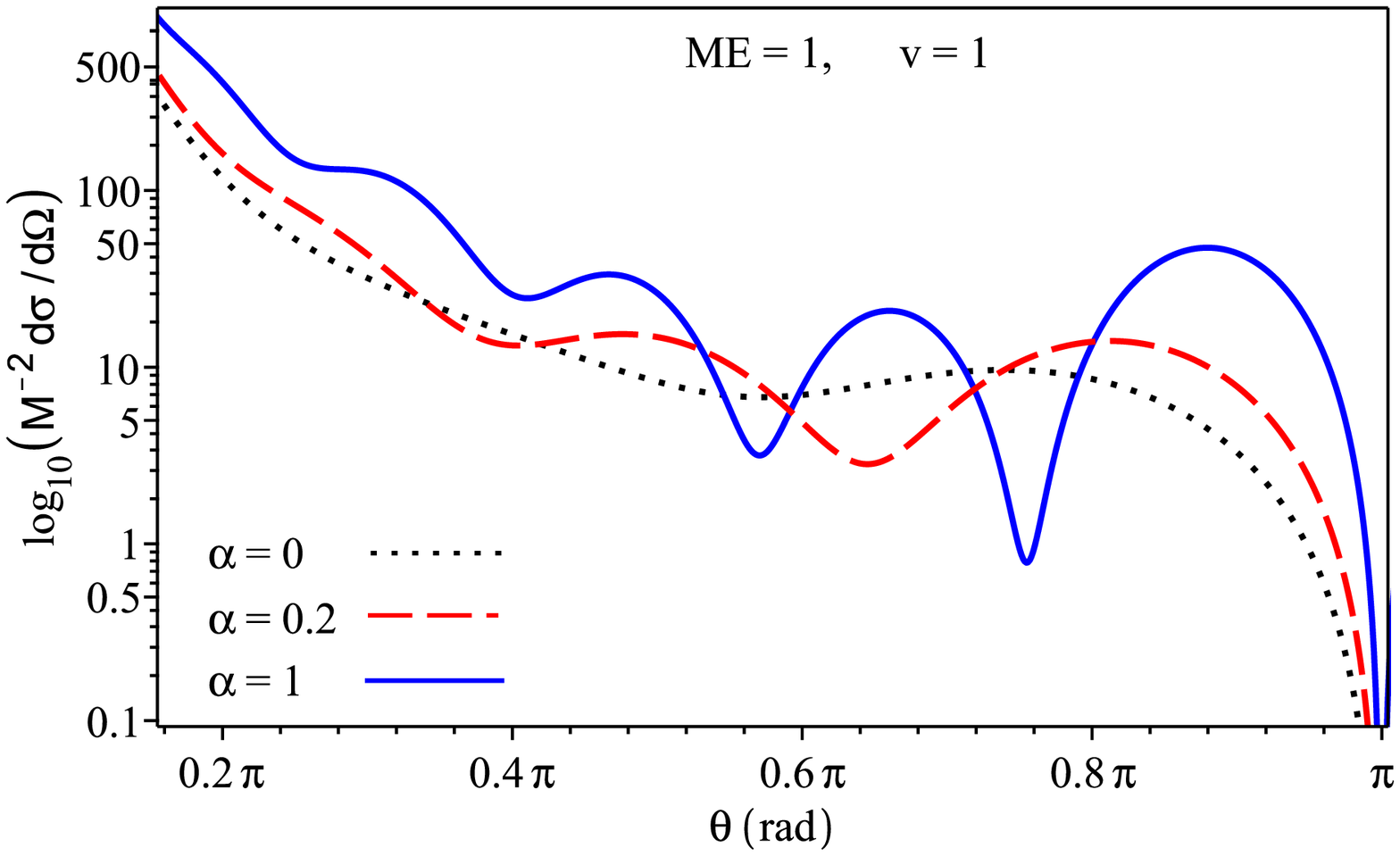}
	\includegraphics[scale=0.3]{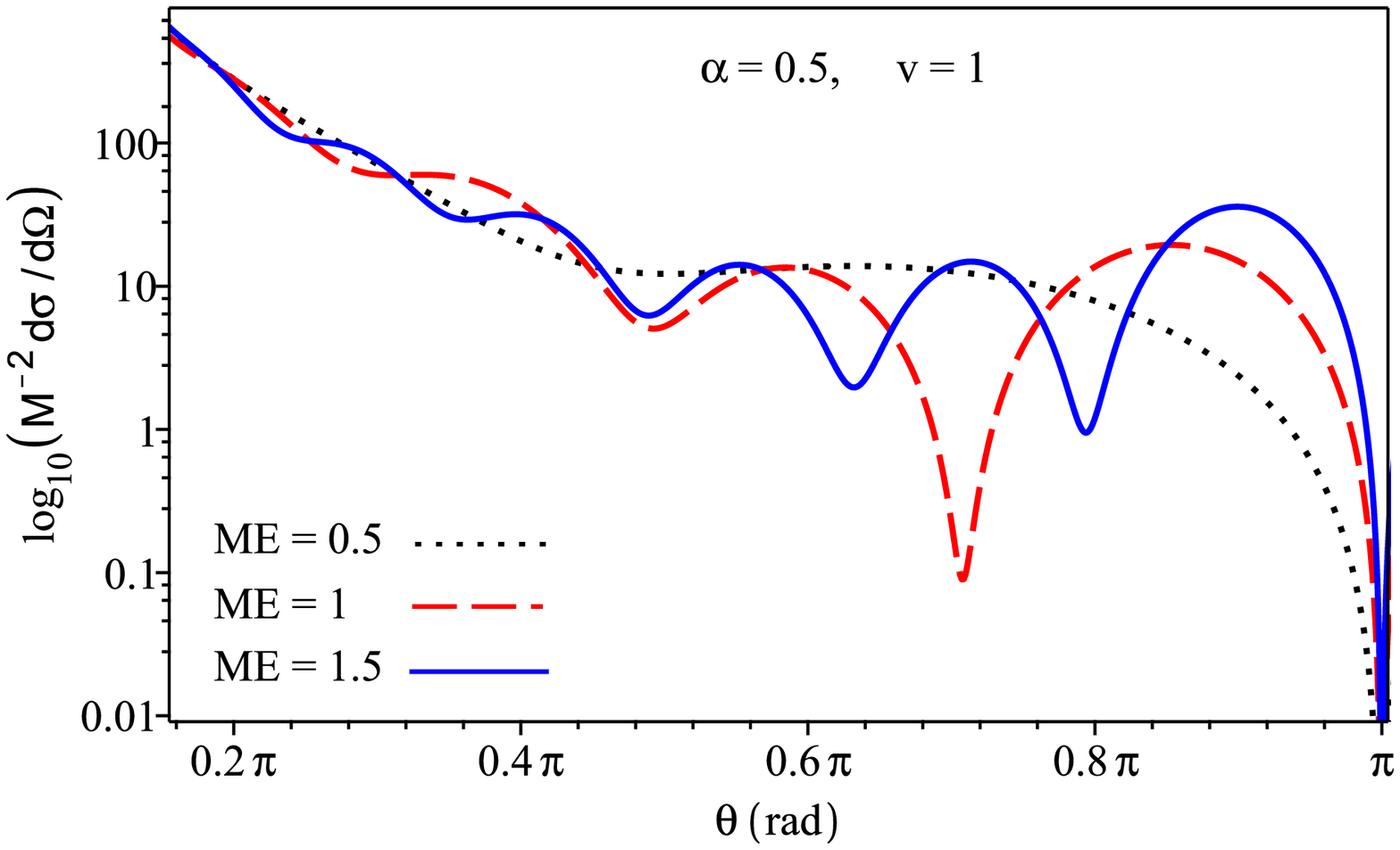}
	\caption{(color online). Scattering of massless fermions by a MOG black hole for fixed $ME=1$ and varying $\alpha = 0, 0.2, 1$ (left panel), respectively fixed $\alpha=0.5$ and varying $ME = 0.5, 1, 1.5$ (right panel). Oscillations start to occur in the scattering intensity as $\alpha$ or $ME$ take higher values and at the same time the glory peak gets narrower.  }
	\label{fig2}
\end{figure*}

In Fig. \ref{fig2} we plot the differential scattering cross section for massless fermions (for which $v=1$, in units of $c$). In the left panel of Fig. \ref{fig2} the frequency is fixed to $ME=1$ and the parameter $\alpha$ takes the values: $\alpha=0$ (the Schwarzschild case) and $\alpha=0.2,\, 1$ (for a MOG black hole), while in the right panel the MOG parameter is set to $\alpha=0.5$ and the frequency takes the values: $ME=0.5,\, 1,\, 1.5$. As can be seen in the right panel, at low frequency ($ME=0.5$) there are no oscillations present in the scattering intensity. However, as the value of $ME$ is increasing, oscillations start to appear, indicating spiral scattering. Moreover, the width of the glory gets narrower for higher values of $ME$. Let us now analyze the left panel of Fig. \ref{fig2}, where we see that for a fixed value of $ME$, one can have spiral scattering by MOG black holes, while in the case of a Schwarzschild black hole ($\alpha=0$) the spiral scattering is absent. Furthermore, the effect of increasing $\alpha$ results in a scattering intensity with a narrower glory width.

\begin{figure*}
	\centering
	\includegraphics[scale=0.3]{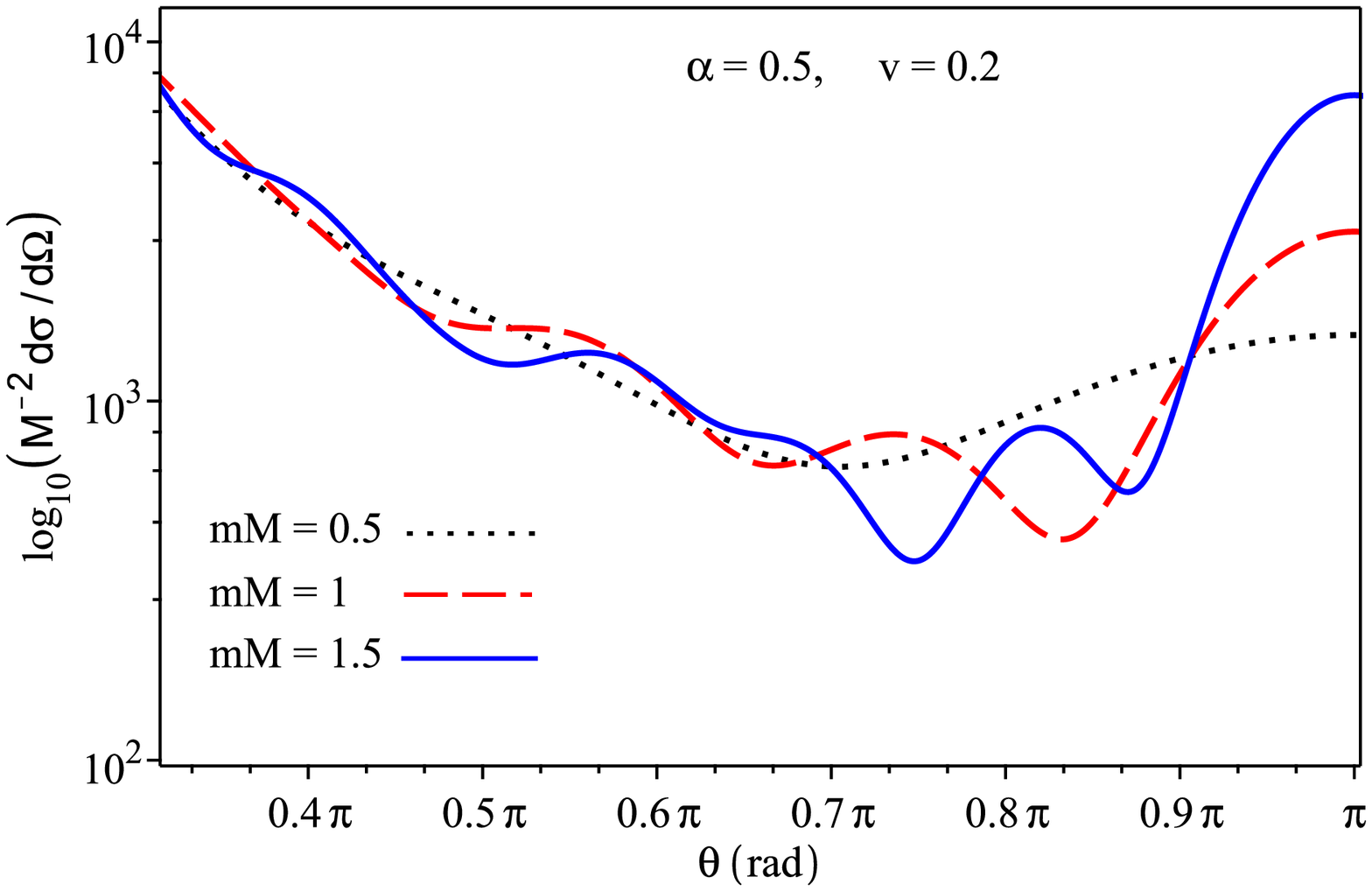}
	\includegraphics[scale=0.3]{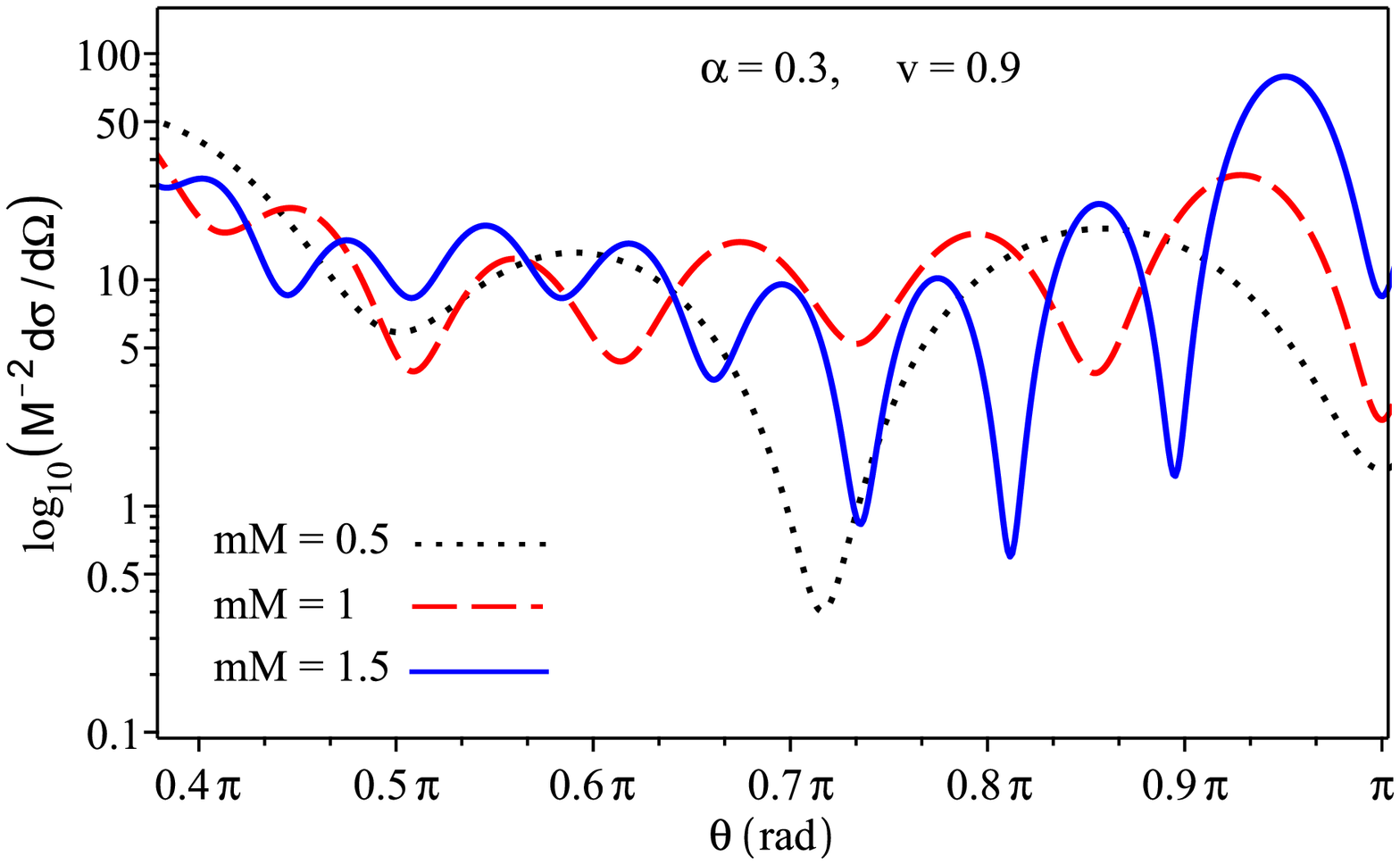}
	\caption{(color online). Comparison between the MOG fermion scattering cross sections at fixed $\alpha=0.5$ and $v=0.2$ (left panel), respectively $\alpha=0.3$ and $v=0.9$ (right panel) for $mM=0.5,\, 1,\, 1.5$. In both cases the glory becomes more significant as the value of $mM$ is increased.  }
	\label{fig3}
\end{figure*}

The plots in Fig. \ref{fig3} are obtained for a fixed value of the MOG parameter $\alpha$ and of the velocity $v$, while the parameter $mM$ takes three different values: 0.5, 1 and 1.5. The parameters $mM$ and $ME$ used to label the figures are not independent, being related by the formula $mM=ME\sqrt{1-v^2}$. Restoring the units one can form the dimensionless parameter $\epsilon=\frac{G_N ME}{\hbar c^3}$ that can be viewed as a measure of the gravitational coupling between the fermion and the black hole. Going back to Fig. \ref{fig3} we observe that the glory peak becomes higher as the value of $mM$ is increased. Furthermore, the oscillations of the orbiting scattering become more significant in the scattering intensity for higher values of $mM$.

\begin{figure*}
	\centering
	\includegraphics[scale=0.3]{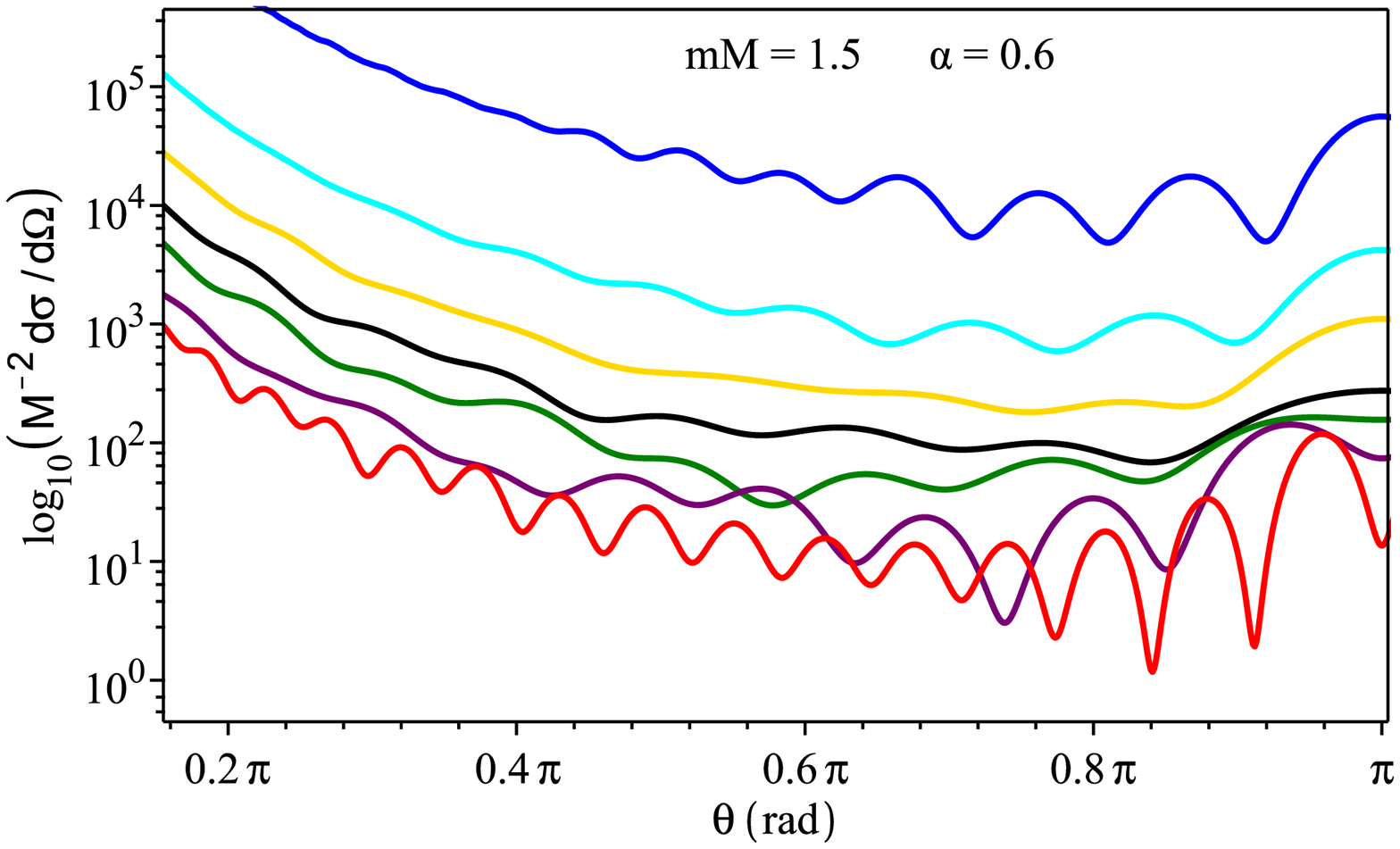}
	\includegraphics[scale=0.3]{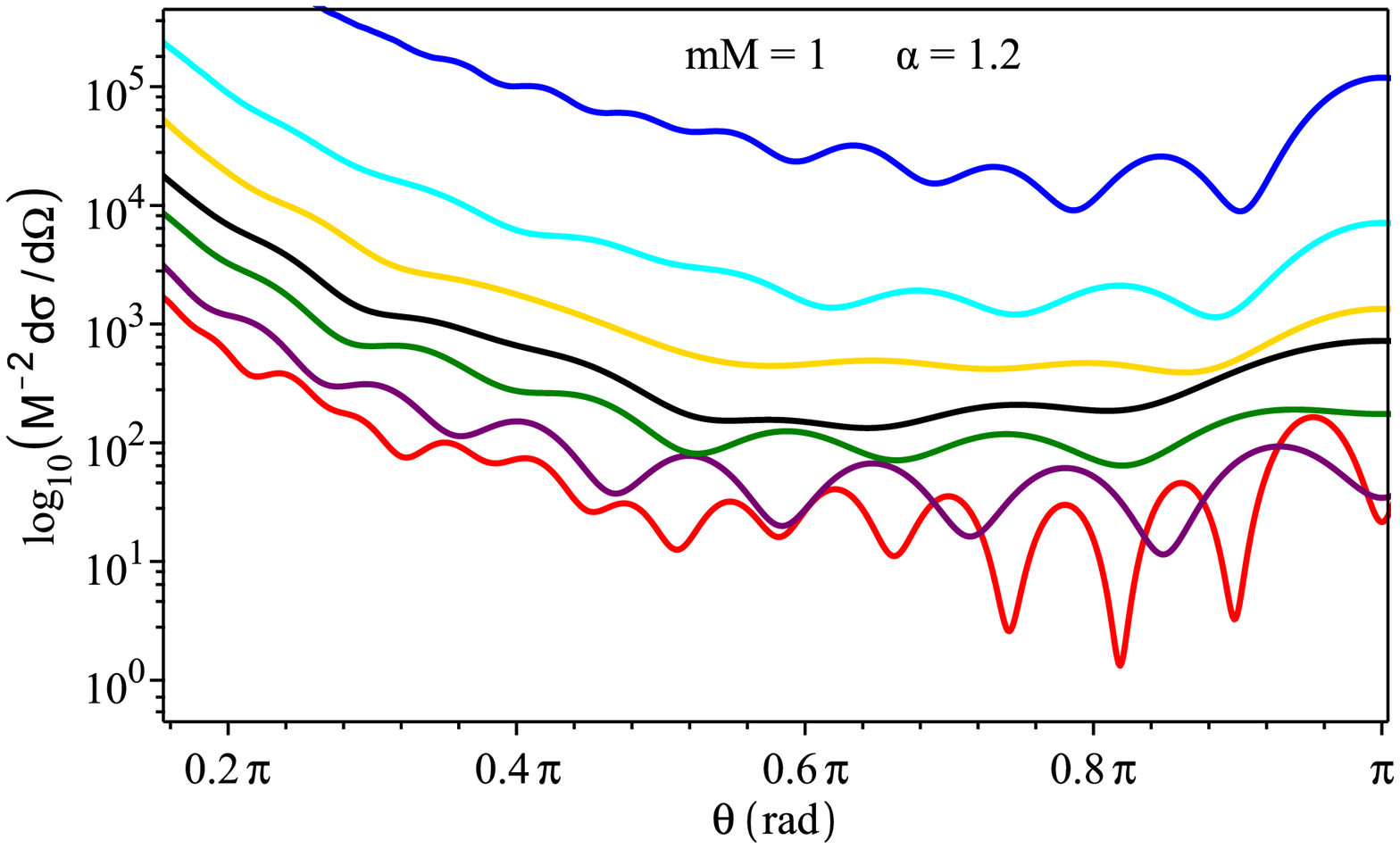}
	\caption{(color online). Scattering intensity for a MOG black hole with $\alpha=0.6$ and $mM=1.5$ (left panel), respectively $\alpha=1.2$ and $mM=1$ (right panel). From top to bottom: v=0.1, 0.2, 0.3, 0.4, 0.5, 0.7, 0.9 in units of $c$. Spiral scattering (signaled by the presence of oscillations in the scattering intensity) is significant mainly for low or high velocities (see blue and red curves) of the incoming fermions. }
	\label{fig4}
\end{figure*}

Fig. \ref{fig4} presents the scattering intensity resulted from the scattering of fermions with a wide range of velocities (top to bottom: $v=0.1, 0.2, 0.3, 0.4, 0.5, 0.7, 0.9 $) by typical MOG black holes. We have set $\alpha=0.6$ and $mM=1.5$ for the plots in the left panel, while for the right panel the values $\alpha=1.2$ and $mM=1$ were used. There are several things to notice in these plots. Let us start with the observation that as the velocity increases from non-relativistic values ($v=0.1$ for top blue curves) to relativistic ones ($v=0.9$ bottom red curves), the scattering intensity passes from having a maxima in the backward direction ($\theta=\pi$) to having a minima. As was shown in Ref. \cite{dolan} for the case of fermion scattering by a Schwarzschild black hole, this effect is due to the interference of spinors that have been rotated in different planes. Furthermore, spiral scattering seems to be more significant for very low or very high values of the velocity $v$, while for intermediate values of $v$ the spiral scattering becomes almost unnoticeable. Also, for scattering of relativistic fermions (bottom red curves) the number of oscillations in the scattering intensity is higher, compared with scattering of non-relativistic fermions (top blue curves). Moreover, for relativistic fermions, spiral scattering can still be observed at low values of the scattering angle, while for non-relativistic fermions, spiral scattering disappears from the scattering intensity at a higher value of $\theta$, as can be seen from comparing the blue and red curves in Fig. \ref{fig4}.

\begin{figure*}
	\centering
	\includegraphics[scale=0.3]{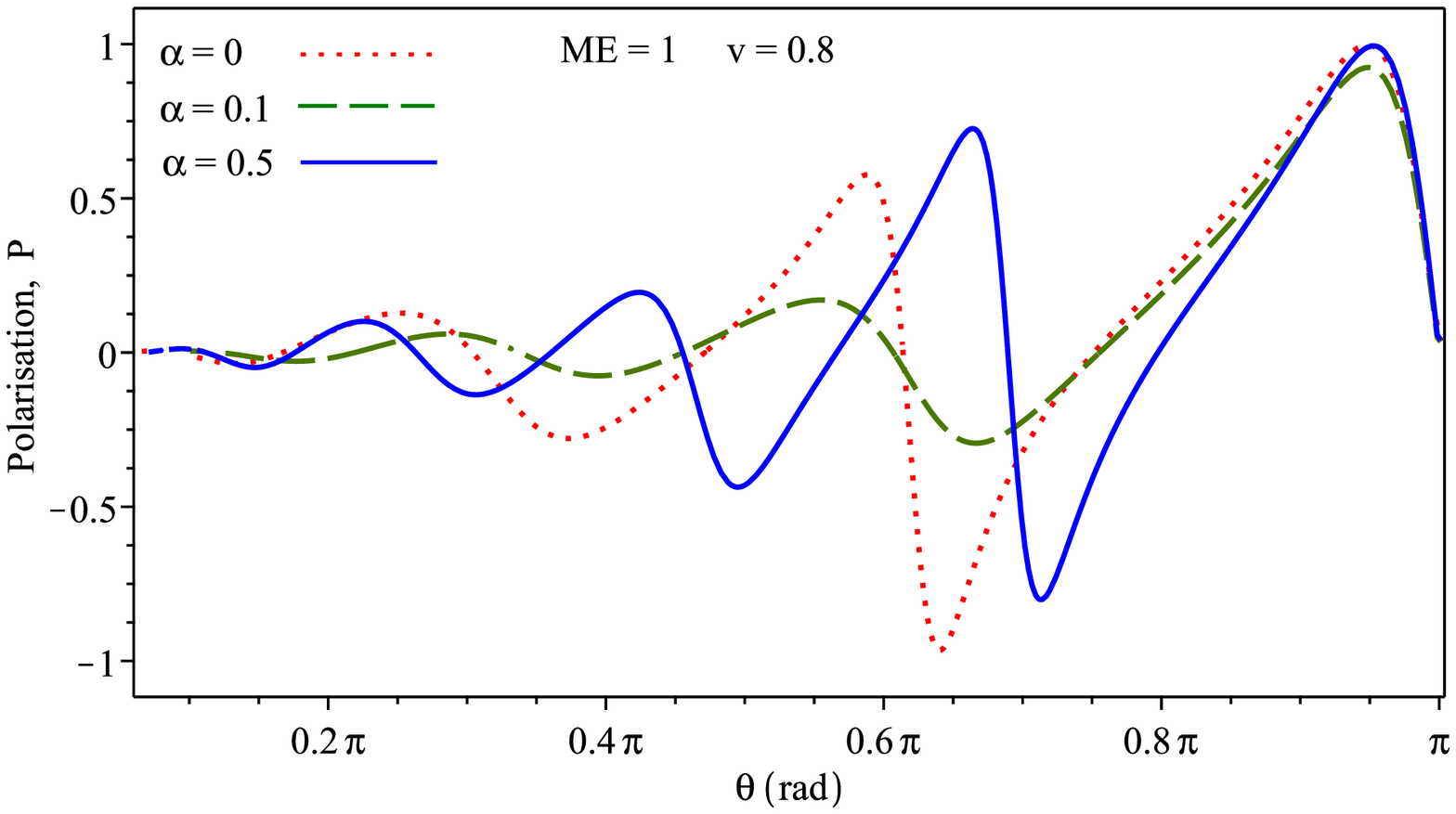}
	\includegraphics[scale=0.3]{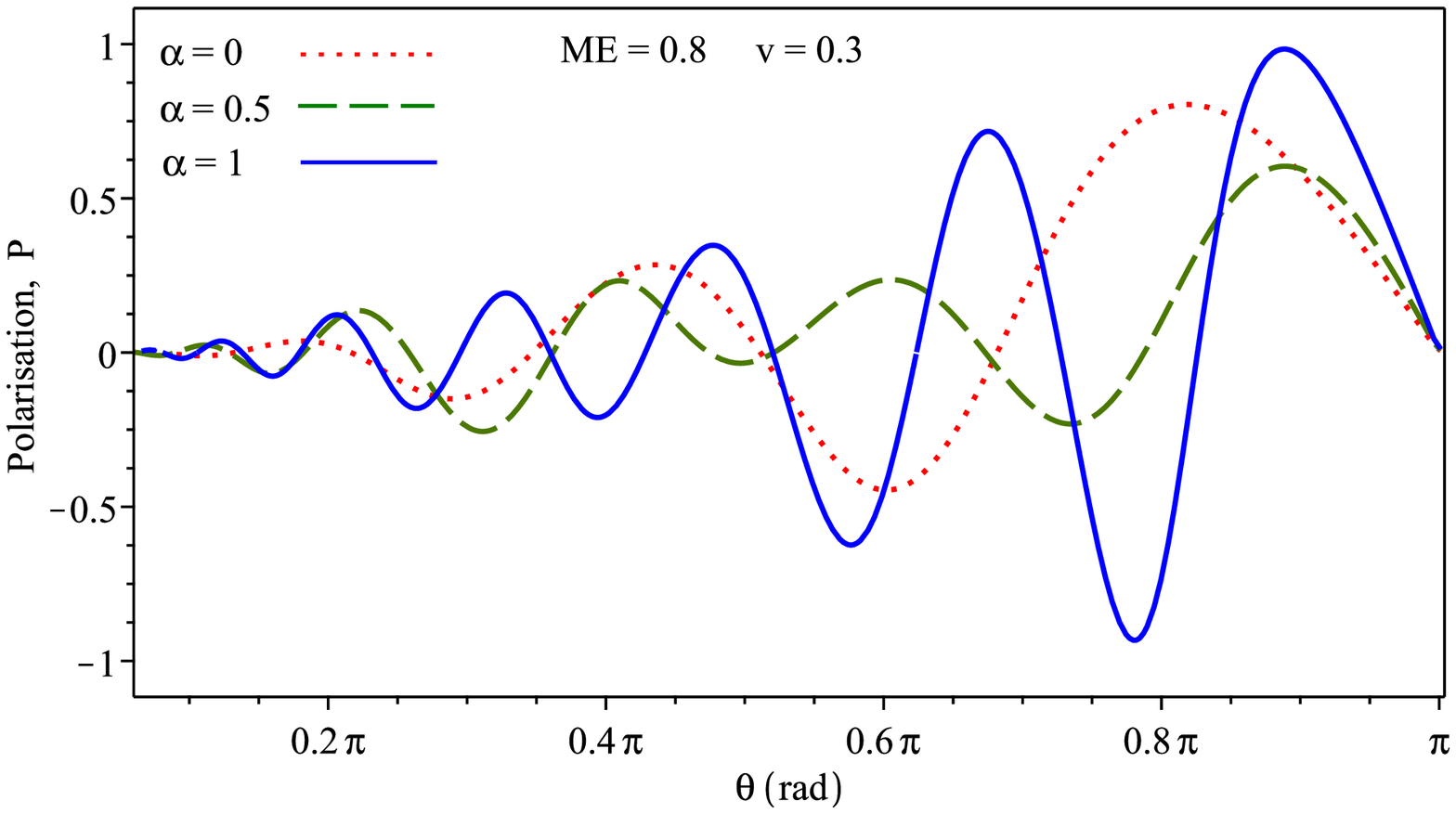}
	\caption{(color online). The induced polarization of fermions scattered by a MOG black hole and comparison with the Schwarzschild case (for which $\alpha=0$). }
	\label{fig5}
\end{figure*}

\begin{figure*}
	\centering
	\includegraphics[scale=0.3]{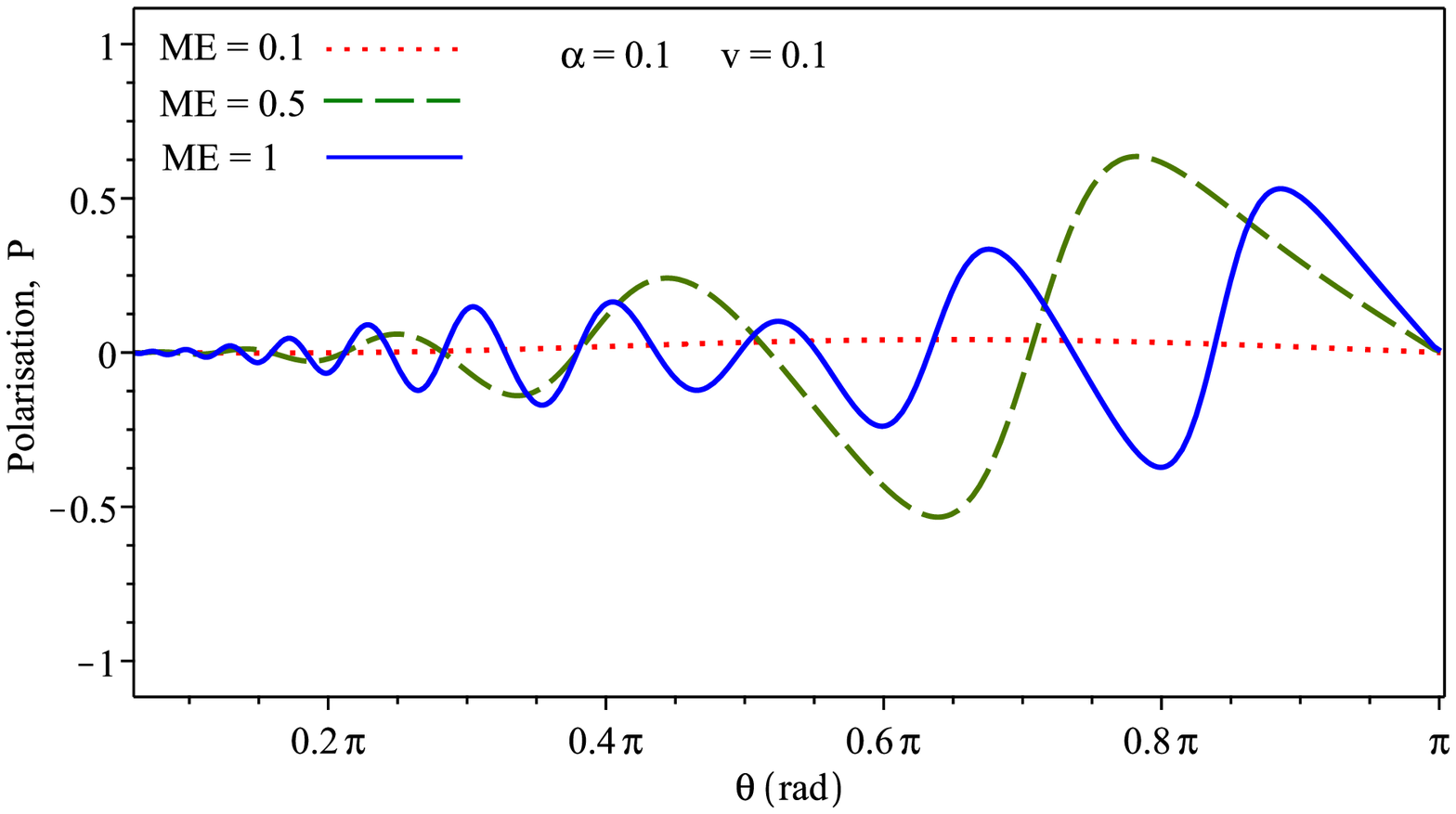}
	\includegraphics[scale=0.3]{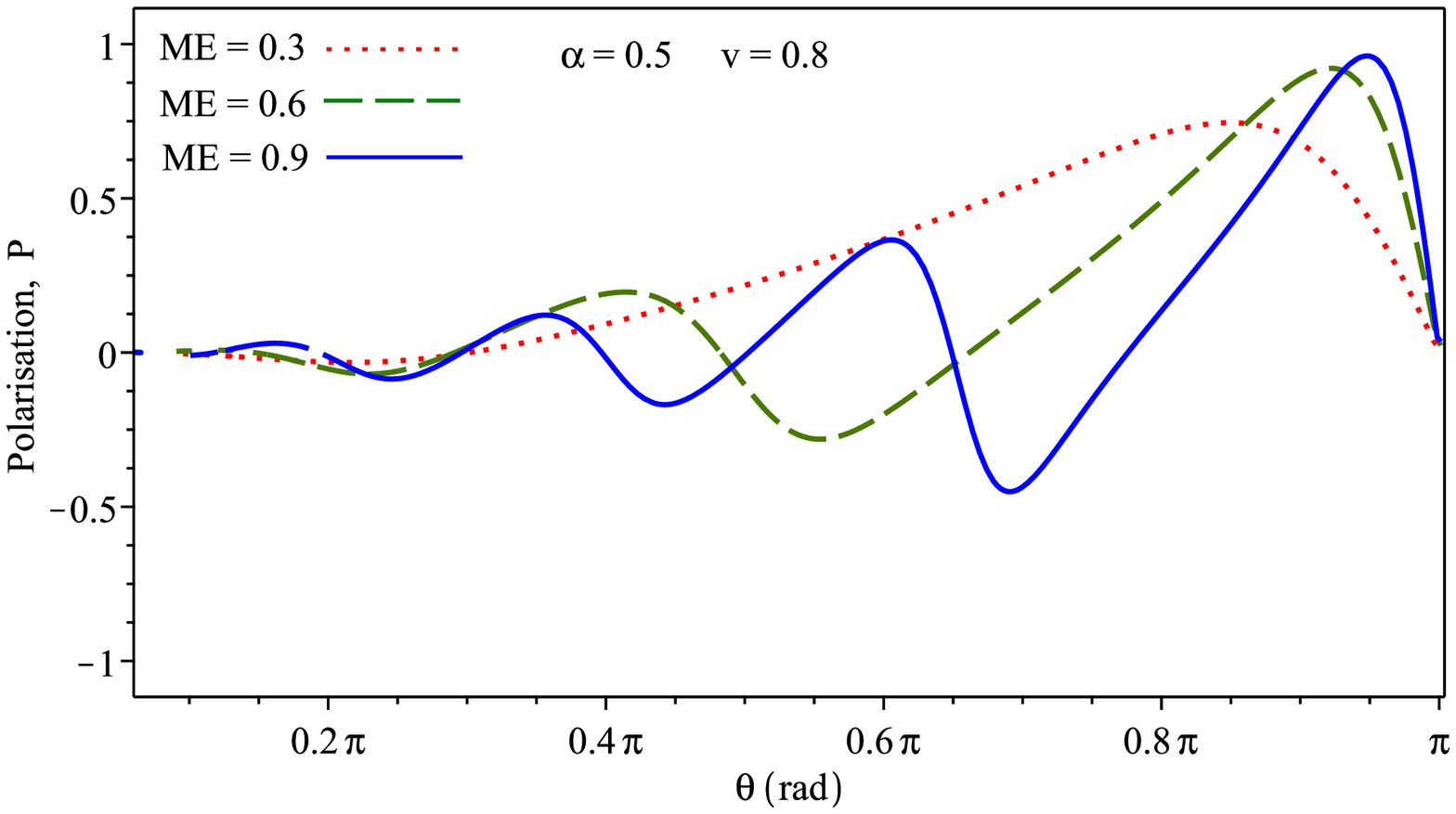}
	\caption{(color online). The induced polarization as function of the scattering angle for a fixed MOG parameter $\alpha$. One can observe that at low $ME$ values there are no oscillations present in the polarization.  }
	\label{fig6}
\end{figure*}

In the case of massive fermions, after the interaction of an initially unpolarized fermion beam with a black hole, the beam could become polarized. The polarization appears due to the rotation of the spin vector after the interaction with the black hole \cite{dolan,Landau,rose}. The plots of the degree of polarization for fermions scattered by a MOG black hole are presented in Fig. \ref{fig5} and Fig. \ref{fig6}. Fixing the frequency $ME$ and the velocity $v$ (Fig. \ref{fig5}), while varying the MOG parameter $\alpha$, will result in a higher number of oscillations present in the polarization, as can be seen from Fig. \ref{fig5}. Thus, one can say that the polarization resulted from fermion scattering by MOG black holes has a more pronounced oscillatory behaviour when compared with Schwarzschild case, for which $\alpha=0$. This can also be correlated with the presence of more oscillations in the scattering intensity (shown in Fig. \ref{fig1}) for values of $\alpha>0$. In Fig. \ref{fig6} the plots of polarization, for fixed MOG parameter $\alpha$ and fixed velocity $v$, are showed. We observe that at very low frequencies (see the curve with $ME=0.1$ in left panel of Fig. \ref{fig6}) the polarization has no oscillations present. As was shown in Ref. \cite{dolan} for the Schwarzschild case, oscillations in the polarization are related to the glory and spiral scattering oscillations, that are absent for very low $ME$ values. Fig. \ref{fig6} indicates that this conclusion remanis valid also for the case of a MOG black hole. However, as we increase the value of the parameter $ME$, oscillations start to occur. The oscillatory behaviour of polarization can be correlated with the existence of glory and spiral scattering, that, as we saw in Figs. \ref{fig1}-\ref{fig4} also induces oscillations in the scattering intensity.

\section{Conclusions}

In this paper we investigated the scattering of fermions by a spherically symmetric black hole coming from a modified gravity (MOG) theory. This type of black hole has a metric whose horizon structure is similar to the Reissner-Nordstrom metric, with the difference that now the charge is of gravitational nature \cite{MOG9} $Q_g=M\sqrt{\alpha G_N}$ and the gravitational constant $G$ differs from Newton's constant $G_N$ by a constant factor $G=G_N(1+\alpha)$.

In Section \ref{sec.results} we analyzed the influence of the MOG parameter $\alpha$ on the scattering intensity $\frac{d\sigma}{d\Omega}$ and the induced polarization $\mathcal{P}$. We saw that higher values of $\alpha$ will produce more oscillations in the scattering intensity (thus accentuating the spiral scattering) and in the induced polarization. In the same time a higher peak in the glory scattering is observed when increasing the value of $\alpha$. This is true for both scattering of massive and massless fermions, as resulted from comparing the scattering cross sections in Figs. \ref{fig1} and \ref{fig2}. Furthermore, when compared with Schwarzschild case (for which $\alpha=0$), scattering by a MOG black hole has an enhanced scattering intensity as can be seen in Figs. \ref{fig1}-\ref{fig2}.

\section*{Acknowledgments}

This work was supported by a grant of Ministery of Research and Innovation, CNCS - UEFISCDI, project number PN-III-P1-1.1-PD-2016-0842, within PNCDI III. I would like to thank Professor Ion I. Cotaescu and to V.E. Ambrus for many fruitful discussions on this topic.

\end{document}